\documentstyle[11pt]{article}

\def \1{\'{\i}}

\def \d{\displaystyle }

\def \pa{ Painlev\'e analysis }

\def \n{\noindent}

\def \smm{singular manifold method }

\def \&{&=&}

\def \t{\paragraph{$\bullet$}}

\def \nn{\nonumber}

\def \q{\quad}

\def \qq{\qquad}

\newcommand{\be}{\begin{equation}}

\newcommand{\ee}{\end{equation}}

\newcommand{\beq}{\begin{eqnarray}}

\newcommand{\eeq}{\end{eqnarray}}

\newcommand{\ba}{\begin{array}}

\newcommand{\ea}{\end{array}}

\begin{document}

\setcounter{equation}{0}

\setcounter{section}{0}

\title{\Large \bf Lax pair, Darboux Transformations and solitonic
solutions for a (2+1) dimensional non-linear Schr\"{o}dinger equation }

\author{ \bf P.G. Est\'evez\footnote{e-mail: pilar@sonia.usal.es } and
G.A. Hern\'aez   \\ {\small
\bf Area de F\1sica Te\'orica}\\ {\small \bf Facultad de F\1sica}\\
{\small \bf Universidad de Salamanca}\\ {\small \bf 37008 Salamanca.
Spain}\\}

\maketitle

\begin{abstract}

 In this paper the Singular Manifold Method 
 has allowed us to obtain the Lax pair, Darboux transformations and
$\tau$ functions for a non-linear Schr\"odiger equation in $2+1$
dimensions. In this  way we can iteratively build  different kind of
solutions with solitonic behavior. 

\end{abstract} \vspace*{0.3in}

\section{Introduction} The integrability and structure of (2+1)
dimensional systems have received considerable attention in the last
few years \cite{ac1992}, \cite{k1993}. Non-linear Schr\"{o}dinger type
equations are a particular case of interest.

These equations were discovered by Calogero \cite{c1975} and then
discussed by Zakharov
\cite{z1980}. Their geometrical properties have been studied in
\cite{ms1992}.

The equation under study in this paper is the following non-linear
Schr\"{o}dinger equation in (2+1)

\beq \nn i\psi_{t} &=& \psi_{xy}+ r^{2}V\psi \\ V_{x} &=&
2\partial_{y}|\psi|^{2} \eeq  Strachan \cite{s1993} rederived this
equation by dimensionally reducing the self-dual Yang Mills equation
\cite{w1977} and proved it to be integrable from the point of view of
geometrical considerations. In \cite{rl1994}  its integrability is
studied in the sense of having the Painlev\'e property and  exact
solutions with solitonic behavior are obtained using Hirota's
bilinear method. One  and two-soliton solutions can be found in
\cite{s1992} and dromionic ones are obtained in
\cite{rl1997} whereas n-soliton solutions may be found in
\cite{mbn1996}.

The geometrical equivalence (called the Lakshmanan equivalence)
between spin systems and non-linear Schr\"{o}dinger equations is
studied in \cite{m1987}, \cite{ml1996}, \cite{md1994} and
\cite{ms1994}.

 Equation (1.1) is the Lakshmanan equivalent of the Myrzakulov-I (M-I)
equation

\beq \nn \vec{S}_t&=&\left(\vec{S}\wedge\vec{S}_y+u\vec{S}\right)_x \\
u_x&=&-\vec{S}\left(\vec{S}_x\wedge\vec{S}_y\right) \eeq proposed in
\cite{m1987} as an extension to (2+1) dimensions of Heisenberg's
1-dimensional spin model \cite{l1977}, \cite{lrt1976}.

The equivalence between (1.1) and (1.2) is proved in \cite{mvsl1998}
and \cite{mvnl1997}.

 If we redefine:

\be \nn \psi=u \qq \psi^{*}=\omega \qq V=-2m_{y} \qq t=it \ee

\n and take $r^{2}=-1$, equation (1.1) becomes:

\beq \nn u_{t}-u_{xy}-2m_{y}u &=& 0 \\ \nn
\omega_{t}+\omega_{xy}+2m_{y}\omega &=& 0 \\ m_{x}+u\omega &=& 0 \eeq
which is the PDE we shall study here.

The plan of this paper is as follows: In section II we shall apply the
\smm to equation (1.4) to obtain the singular manifold equations. In
section 3 we use the SMM to linearize the singular manifold equations
and obtain the Lax pair. Section 4 is devoted to determining the
Darboux transformations and $\tau$-functions. We apply the results of
section 4 to obtain solitonic solutions in section 5. The conclusions
are presented in section 6.

\section{Singular Manifold Method}

\setcounter{equation}{0}

\subsection*{Leading term analysis} In order to perform the \pa
\cite{pp} for equation (1.4) we need to expand the fields $u$, $\omega$
and $m$ in a generalized Laurent expansion  in terms of an arbitrary
singularity manifold $\chi(x,y,t)=0$. Such an expansion should be of
the form \cite{WTC}:
\beq \nn u &=& \sum_{j=0}^{\infty} u_j(x,y,t)\,
[\chi(x,y,t)]\,^{j-\alpha}\\ \nn \omega  &=&
\sum_{j=0}^{\infty}
\omega_j(x,y,t) \, [\chi(x,y,t)]\,^{j-\beta} \\
m&=&\sum_{j=0}^{\infty} m_j(x,y,t)\, [\chi(x,y,t)]\,^{j-\gamma} \eeq
By substituting (2.1) in (1.4), we have for the leading terms:
\be \alpha=\beta=\gamma=1 \qq m_{0}=\chi_{x} \qq
u_{0}\,\omega_{0}=\chi_x^2 \ee from which we see that leading analysis
is not able to determine $u_0$ and $\omega_0$ independently and only
gives us their product. This suggests that we should  write the
dominant terms $u_0$ and $\omega_0$ in the more general way as:
\be  u_0=A(x,y,t)\chi_x \qq \omega_0={1\over A}\chi_x \ee
\subsection*{Truncated expansions. Auto-B\"{a}cklund transformations}
If we truncate  expansions (2.1) at the constant level, as  is
required by the SMM
\cite{Weiss}, we can write the solutions in terms of a singular
manifold, $\phi$, which is not yet an arbitrary function because it is
determined by the truncation condition. We can therefore write the
solutions (2.1) to equation (1.4) in the following way:
\beq \nn m'&=&m+{\phi_x\over\phi} \\ \nn u'&=&u+{{A\phi_x}\over\phi} \\
\omega'&=&\omega+{\phi_x\over A\phi} \eeq The set of equations (2.4)
are the auto-B\"{a}cklund transformations between two solutions of
(1.4).
\subsection*{Expression of the solutions in terms of the Singular
Manifold} Substituting  equations (2.4) in (1.4), we obtain a
polynomial in $\phi$. If we require all the coefficients of this
polynomial to be zero we obtain the following expressions after some
algebraic manipulations (we used MAPLE V to handle the calculation.
The details are in the appendix):
\beq u&=& -{A\over2}\left (v+ \left ({A_x\over A}+h\right )\right) \\
\omega &=& -{1\over {2A}}\left(v-\left({A_x\over A}+h\right)\right) \\
m_x&=&{1\over 4}\left(\left({A_x\over A}+h\right)^2-v^2 \right) \\
m_y&=&{1\over 2}\left({A_t\over A}-{A_x\over A}{A_y\over A}-v_y \right
)\eeq where $v$, $w$ and $q$ are defined as:
\beq \nn v&=&{\phi_{xx}\over \phi_x} \\ w&=&{\phi_t\over\phi_x}\\ \nn
q&=&{\phi_y\over\phi_x} \eeq   and $h=h(y,t)$  is an $x$-independent
function  which arises after performing an integration in
$x$ (see  appendix).

It is useful to notice that the compatibility conditions between the
definitions (2.9) give rise to the following equations:

\beq \nn \phi_{xxt}=\phi_{txx} &\Longrightarrow& v_t=(w_x+vw)_x \\
\phi_{xxy}=\phi_{yxx}&\Longrightarrow& v_y=(q_x+vq)_x\\ \nn
\phi_{yt}=\phi_{ty} &\Longrightarrow& q_t=w_y+wq_x-qw_x \eeq

\subsection*{Singular manifold equations} Furthermore,  substitution
of (2.4) in (1.4) provides equations to be satisfied by the singular
manifold. These equations are (see  appendix):
\beq 0&=&w+hq-{A_y\over A}\\0&=&
\left({A_xA_y\over A^2}-{A_t\over A}\right)_x+\left(v_x-{v^2\over
2}+{1\over 2}\left({A_x\over A}+h\right)^2\right)_y\eeq and the
following equation for $h$

\be h_t+hh_y=0 \ee The set (2.10)-(2.13) are the singular manifold
equations.

\section{Lax pairs}

\setcounter{equation}{0}

\subsection{Painlev\'e analysis in singular manifold equations}

We can consider the singular manifold equations (2.10)-(2.12) as a
system of non-linear coupled PDE's in $v$, $q$, $w$
 and $A$.  It is useful to define
\beq \nn \alpha&=&{A_x\over A}+h\\ \nn \beta&=&{A_y\over A}+h_yx\\ 
\gamma&=&{A_t\over A}+h_tx\eeq in which case (2.10)-(2.12) can be
combined to give:
\beq  &0&= v_t+hv_y-[(\beta-h_yx)_x+(\beta-h_yx) v]_x\\ &0&=
\left(\beta\alpha-h\beta-x\alpha
h_y-\gamma\right)_x+\left(v_x-{v^2\over 2}+ {\alpha^2\over
2}\right)_y\eeq This allows us to perform the leading terms analysis
by setting:

\beq \nn v &\sim& v_0\chi^a \\ \alpha &\sim & \alpha_0\chi^b, \q \beta
\sim   \beta_0\chi^b, \q
\gamma \sim  \gamma_0\chi^b\eeq Substitution of (3.4) in (3.2)-(3-3) 
yields the leading powers:
\be a=b=-1  \ee and the leading coefficients:
\beq \nn v_0&=&\chi_x \\ \alpha_0&=&\pm \chi_x, \q \beta_{0}=\pm
\chi_y, \q \gamma_{0}=\pm
\chi_t
\eeq   The $\pm$ sign  tells us that the Painlev\'e expansion has two
branches. The problem of systems with two Painlev\'e branches has been
discussed in \cite{CMP}, \cite{EG}, \cite{EG2},
\cite{ECG}. These references suggest that we should consider both
branches  simultaneously  by using two singular manifolds, one for
each branch.

\subsection{Eigenfunctions and the singular manifold}

In agreement with the foregoing, we can write the dominant terms of
$v$ and $A$ as:

\beq \nn v &=& {\psi_x^+\over\psi^+}+{\psi_x^-\over\psi^-}\\  \nn
\alpha &=& {\psi_x^+\over\psi^+}-{\psi_x^-\over\psi^-}, \qquad
\quad\beta = {\psi_y^+\over\psi^+}-{\psi_y^-\over\psi^-},\qquad
\qquad\q  \gamma =
{\psi_t^+\over\psi^+}-{\psi_t^-\over\psi^-}\\{A_x\over A} &=&
{\psi_x^+\over\psi^+}-{\psi_x^-\over\psi^-}-h, \q{A_y\over A} =
{\psi_y^+\over\psi^+}-{\psi_y^-\over\psi^-}-h_yx,\q {A_t\over A} =
{\psi_t^+\over\psi^+}-{\psi_t^-\over\psi^-}-h_tx\eeq where $\psi^+$ is
the singular manifold for the positive branch and $\psi^-$ for the
negative one. We will see later on that $\psi^+$ and $\psi^-$ are the
eigenfunctions of the Lax pair.

Integrating (3.7), we obtain the expressions of $\phi$ and $A$ in terms
of the eigenfunctions:

\beq \nn \phi_x&=&\psi^+\psi^-, \quad\qquad
\phi_t+h\phi_y=\psi^-\psi^+_y-\psi^+\psi^-_y-h_yx\psi^+\psi^-\\ A &=&
{\psi^+\over\psi^-}e^{-hx}
\eeq

\subsection{Linearization of the singular manifold equations: the Lax
pair}

Substitution of (3.8) in (2.5)-(2.8) gives us the following expressions
of
$u$,
$\omega$, $m_x$ and $m_y$ in terms of $\psi^+$ and 

\beq u &=& -{\psi_x^+\over \psi^-}e^{-hx} \\ \omega &=&
-{\psi_x^-\over \psi^+}e^{hx} \\ m_x &=&
-{{\psi_x^+\psi_x^-}\over{\psi^+\psi^-}} \\ m_y &=& 
\left({\psi_t^++h\psi_y^++u_ye^{hx}\psi^-\over 2\psi^+}\right)+
\left({-\psi_t^--h\psi_y^-+w_ye^{-hx}\psi^+\over 2\psi^-}\right)
\eeq  Equations (3.9) and (3.10) can be considered to be the spatial
part of the Lax pair, which written in a more apropiate way reads:

\beq \nn 0&=&\psi_x^+ + u\psi^-e^{hx}  \\
0&=&\psi_x^-+\omega\psi^+e^{-hx}  \eeq

The temporal part of the Lax pair can be obtained as follows: If we 
substitute (3.7) in (3.2) and use (3.13),  we obtain  (after an
integration in
$x$)the equation:

\be
\left({\psi_t^++h\psi_y^++u_ye^{hx}\psi^-\over -\psi^+}\right)-
\left({-\psi_t^--h\psi_y^-+w_ye^{-hx}\psi^+\over -\psi^-}\right)-h_y=0
\ee Adding and subtracting (3.12) and (3.14) we obtain  the temporal
part of the Lax pair:

\beq \nn \psi_t^+ - m_y\psi^+ + h\psi_y^+ +
u_y\psi^-e^{hx}+{1\over2}h_y\psi^+ &=& 0\\ \psi_t^- + m_y\psi^- +
h\psi_y^- - \omega_y\psi^+e^{-hx} + {1\over2}h_y\psi^-&=& 0 \eeq  It is
interesting to notice that the compatibility condition between (3.13)
and (3.15) is the equation (1.4) together with the condition (2.13).
Therefore (3.13) and (3.15) form  the the Lax pair for (1.4) and $h$
is the spectral parameter although it is {\bf non-isospectral}.

\section{Darboux transformations}

\setcounter{equation}{0}

Summarizing the above results:

\t Let be $u$, $\omega$ and $m$ solutions of (1.4) and $\phi_1$ a
singular manifold for them. This singular manifold can be constructed
from two eigenfunctions
$\psi_1^+$ and $\psi_1^-$ as:

\be \phi_{1x}=\psi_1^+\psi_1^- \ee where $\psi_1^+$ and $\psi_1^-$
satisfy the Lax pairs:

 \be  \ba{ll}  0=\psi_{1x}^+ + u\psi_1^-e^{h_1x}  \\ 
0=\psi_{1x}^-+\omega\psi_1^+e^{-h_1x} \\   0=\psi_{1t}^+ - m_y\psi_1^+
+ h_1\psi_{1y}^+ + u_y\psi_1^-e^{h_1x}+{1\over2}h_{1y}\psi_1^+ \\
0=\psi_{1t}^- + m_y\psi_1^- + h_1\psi_{1y}^- -
\omega_y\psi_1^+e^{-h_1x} + {1\over2}h_{1y}\psi_1^- 
\ea \ee and the spectral parameter $h_1$ satisfies:
\be h_{1t}+h_1h_{1y}=0 \ee

\t Substituting  (3.8) in (2.4) , we can define new solutions $u'$,
$\omega'$ and $m'$:

\beq \nn u'&=& u + {(\psi_1^+)^{2}e^{-h_1x}\over\phi_1} \\ \nn \omega'
&=& \omega+ {(\psi_1^-)^{2}e^{h_1x}\over\phi_1} \\ m' &=&
m+{\phi_{1x}\over\phi_1} \eeq whose Lax pairs will be:

 \be  \ba{ll}  0=\psi_x^{'+} + u'\psi^{'-}e^{h_2x}  \\ 
0=\psi_x^{'-}+\omega'\psi^{'+}e^{-h_2x} \\   0=\psi_t^{'+} -
m'_y\psi^{'+} + h_2\psi_y^{'+} +
u'_y\psi^{'-}e^{h_2x}+{1\over2}h_{2y}\psi^{'+} \\ 0=\psi_t^{'-} +
m'_y\psi^{'-} + h_2\psi_y^{'-} - \omega'_y\psi^{'+}e^{-h_2x} +
{1\over2}h_{2y}\psi^{'-}  \ea \ee  and we can construct a singular
manifold $\phi'$ for the iterated fields $u'$, $\omega'$ and
$m'$ through
$\psi^{'+}$ and
$\psi^{'-}$ as:

\be \phi'_x=\psi^{'+}\psi^{'-} \ee

\subsection*{Truncated expansion in the Lax pair}

We can consider the Lax pair (4.5) as a system of coupled non-linear
PDE's (\cite{EG2}, \cite{KS})  in  $\psi^{'+}$, $\psi^{'-}$, $m'$,
$u'$ and $\omega'$. Therefore, the singular manifold method can be
applied to the Lax pair itself and  truncated expansions for
$\psi^{'+}$ and
$\psi^{'-}$ should be added to the expansions (4.4). Such expansions
can be written as:

\be \psi^{'+}=\psi_2^+ - {{\psi_1^+\Omega^+}\over\phi_1} \qq
\psi^{'-}= \psi_2^- - {{\psi_1^-\Omega^-}\over\phi_1} \ee 

The seminal solutions $m$, $u$, $\omega$, $\psi_2^+$ and $\psi_2^-$
must satisfy the same Lax pair with the spectral parameter $h_2$,
which means:

 \be  \ba{ll}  0=\psi_{2x}^+ + u\psi_2^-e^{h_2x} \\ 
0=\psi_{2x}^-+\omega\psi_2^+e^{-h_2x} \\   0=\psi_{2t}^+ - m_y\psi_2^+
+ h_2\psi_{2y}^+ + u_y\psi_2^-e^{h_2x}+{1\over2}h_{2y}\psi_2^+ \\ 
0=\psi_{2t}^- + m_y\psi_2^- + h_2\psi_{2y}^- -
\omega_y\psi_2^+e^{-h_2x} + {1\over2}h_{2y}\psi_2^- 
\ea \ee

Substituting the truncated expansions (4.4) and (4.7) in the Lax pair
(4.5) and after some calculation (we used MAPLE V for it) we obtain:

\beq  \nn \Omega^+ &=&
{{\psi_1^+\psi_2^-e^{-(h_1-h_2)x}-\psi_2^+\psi_1^-}\over{h_2-h_1}} \\
\Omega^- &=& {{\psi_1^+\psi_2^- -
\psi_2^+\psi_1^-e^{(h_1-h_2)x}}\over{h_2-h_1}} \eeq

Summarizing: the set of equations

\beq \nn u'&=& u + {{\psi_1^{+^{2}}e^{-h_1x}}\over\phi_1} \\ \nn
\omega' &=& \omega+ {{\psi_1^{-^{2}}e^{h_1x}}\over\phi_1} \\ \nn m'
&=& m+{\phi_{1x}\over\phi_1} \\ \nn
\psi^{'+}&=&\psi_2^+ - {{\psi_1^+\Omega^+}\over\phi_1} \\ \psi^{'-}
&=& \psi_2^- - {{\psi_1^-\Omega^-}\over\phi_1} \eeq where $\Omega^+$
and $\Omega^-$ are given by (4.9), constitutes a transformation of
potentials and eigenfunctions that  leaves the Lax pairs invariant.
Hence, (4.10) should be considered as a Darboux transformation
\cite{MS}.

\section{Iteration of the singular manifold: $\tau$-functions}

\setcounter{equation}{0}

The $\tau$-functions of Hirota's bilinear method \cite{Hirota} can be
built through the singular manifold and Darboux transformations as
follows:

Equation (4.6) can be considered as a non-linear equation in $\phi'$,
$\psi^{'+}$ and $\psi^{'-}$ and it is therefore pertinent to add  the
following truncated expansion to the set (4.10):
\be \phi'= \phi_2+{\Delta\over\phi_1} \ee where $\phi_2$ satisfies:
\be \phi_{2x}=\psi_2^+\psi_2^- \ee Substituting (5.1) and (4.7) in
(4.6) one has:
\be \Delta=-\Omega^+\Omega^- \ee Since (5.1) defines a singular
manifold for $m'$, can be used  to build an iterated solution:
\be m''=m'+{\phi'_x\over\phi'} \ee Substitution of equation (4.4) for
$m'$ in (5.4) gives:
\be m''=m+{\tau_x\over\tau} \ee where
\be \tau=\phi'\phi_1=\phi_1\phi_2-\Omega^+\Omega^- \ee is the Hirota
$\tau$-function.

\section{Solutions}

\setcounter{equation}{0}

In this section, we  obtain  solutions to the system (1.4) in a
systematic way using the previous results. The steps followed in this
iterative procedure can be summarized as:

\t 1) We start from seminal solutions of (1.4) and write the Lax pair
for them.The solitonic or dromionic behavior of the iterated
solutions will depend on our choice of the seminal ones.

\t 2) Solving the Lax pairs, we obtain $\psi_1^+$, $\psi_1^-$,
$\psi_2^+$ and $\psi_2^-$.

\t 3) We use the results of 2) in (4.1), (4.9), and (5.2) to obtain
$\phi_1$, $\phi_2$, $\Omega^+$ and
$\Omega^-$.

\t 4) We use (4.4) and (5.4) to obtain the first and second iterations
$m'$ and $m''$, respectively.

\subsection{Line solitons $m=-abx$, $u=a$, $\omega=b$}

If we restrict ourselves to the case in which  $h_1$ and $h_2$ are
constants, non-trivial solutions of the Lax pairs (4.2) and (4.8) are:

\beq \nn
\psi_1^+=\exp\left[\alpha_1x+\beta_1y+\left({{ab}\over\alpha_1}-\alpha_1\right)\beta_1t\right]
&
\psi_1^-=-{\alpha_1\over a}\exp\left[{{ab}\over\alpha_1}x+\beta_1y+
\left({{ab}\over\alpha_1}-\alpha_1 \right)\beta_1t\right] \\
\psi_2^+=\exp\left[\alpha_2x+\beta_2y+\left({{ab}\over\alpha_2}-\alpha_2\right)\beta_2t\right]
&
\psi_2^-=-{\alpha_2\over a}\exp\left[{{ab}\over\alpha_2}x+\beta_2y+
\left({{ab}\over\alpha_2}-\alpha_2 \right)\beta_2t\right] \eeq where 
$\beta_1$ $\beta_2$ are arbitrary constants and $\alpha_1$ and
$\alpha_2$  are related to the spectral parameter as:
\be h_i=\alpha_i-{{ab}\over\alpha_i} \ee

If we define: 

\beq \nn
P_1=\exp\left[\alpha_1x+\beta_1y+\left({{ab}\over\alpha_1}-\alpha_1\right)\beta_1t\right]
& Q_1=\exp\left[{{ab}\over\alpha_1}x+\beta_1y+
\left({{ab}\over\alpha_1}-\alpha_1
\right)\beta_1t\right] \\
P_2=\exp\left[\alpha_2x+\beta_2y+\left({{ab}\over\alpha_2}-\alpha_2\right)\beta_2t\right]
& Q_2=\exp\left[{{ab}\over\alpha_2}x+\beta_2y+
\left({{ab}\over\alpha_2}-\alpha_2
\right)\beta_2t\right] \eeq integration of (3.8) yields:

\beq \nn \phi_1=-{\alpha_1\over
a}{1\over{\alpha_1+{{ab}\over\alpha_1}}}(c_1+P_1Q_1) \\
\phi_2=-{\alpha_2\over
a}{1\over{\alpha_2+{{ab}\over\alpha_2}}}(c_2+P_2Q_2) \eeq where $c_1$
and $c_2$ are arbitrary constants. Using (4.9) one has:

\beq \nn
\Omega^+=-{{\alpha_1\alpha_2}\over{a(\alpha_1\alpha_2+ab)}}P_2Q_1 \\ 
\Omega^-=-{{\alpha_1\alpha_2}\over{a(\alpha_1\alpha_2+ab)}}P_1Q_2 \eeq
The first iteration provides the solution (figure 1):

\be m'_x= -ab+\partial_{xx}[ln\phi_1] \ee and the second (figure 2):

\be m''_x=-ab+\partial_{xx}[ln\tau] \ee where

 \be \phi_i=-{\alpha_i\over
a}{c_i\over{\alpha_i+{{ab}\over\alpha_i}}}(1+F_i) \ee

\be \tau=\phi_1\phi_2 -\Omega^+\Omega^-={{\alpha_1\alpha_2}\over
a^2}{1\over{\left(\alpha_1+{{ab}\over\alpha_1}\right)\left(\alpha_2+{{ab}\over\alpha_2}\right)}}[1+F_1+F_2-A_{12}F_1F_2]
\ee and 

\be
F_i(x,y,t)=\exp\left[\left(\alpha_i+{{ab}\over\alpha_i}\right)x+2\beta_i
y-2\left(\alpha_i-{{ab}\over \alpha_i}\right)\beta_i t
+\varphi_i\right] \ee

\be A_{12}={{ab(\alpha_1-\alpha_2)^2}\over{(\alpha_1\alpha_2+ab)^2}}
\ee and we have redefined $c_i$ as: $c_i=e^{-\varphi_i}$.

A particularly  interesting case occurs when $\alpha_1=\alpha_2$ and
hence the interaction term
$A_{12}$ vanishes. This case is termed the {\it resonant state}
\cite{EL} (figure 3).

\subsection {Dromions $m=0$, $u=0$, $\omega=b$}

For this seminal solution (similar solutions are obtained for $u=a$,
$\omega=0$), assuming that $h_1$
 and $h_2$ are constants, the easiest non-trivial solutions of (4.2)
and (4.8) are:

\beq \nn \psi_1^+=K_1(y,t) & \q \psi_1^-={b\over h_1}e^{-h_1x}K_1(y,t)
\\ \psi_2^+=K_2(y,t) &
\q \psi_2^-={b\over h_2}e^{-h_2x}K_2(y,t)
\eeq where $K_i$ are $x$-independent functions that satisfy
\be K_{it}+h_iK_{iy}=0\ee Integration of (3.8) yields:

\be \phi_i=-{b\over h_i^2} \left( R_i(y,t) +
K_i^2(y,t)e^{-h_ix}\right) \ee where \be R_{it}+h_iR_{iy}=0\ee and
from (4.9):
\beq \nn \Omega^+&=& -{b\over h_1h_2}K_1K_2e^{-h_1x}\\ \Omega^+&=&\
-{b\over h_1h_2}K_1K_2e^{-h_2x}\eeq

\n The first iteration provides :
\be m'_y=\partial_{xy}\ln\, [\phi_1] \ee and the second one 
\be m''_y=\partial_{xy}\ln\,[\tau]\ee where
\be \tau=\phi_1\phi_2-\Omega^+\Omega^-={b^2\over
h_1^2h_2^2}\left\{R_1R_2+R_2K_1^2e^{-h_1x}+R_1K_2^2e^{-h_2x}\right\}\ee
Due to the arbitrariness of $R_i$ and $K_i$, (6.17) and (6.18) are a
rich collection of solutions. Let us consider  some particular cases

\subsubsection*{Case 1} This  corresponds to the choice: $$ R_i=
1+{\displaystyle e^{ci(y-h_it)}}$$ $${\d K^2_i= 1+a_ie^{c_i(y-h_it)}}$$
Fig. 4 represents $m'_y$ for this choice. Figures 5.a-5.c are the
second iteration $m''_y$ for different times.

\subsubsection*{Case 2} Another possibility is:$$  R_1=
1+{\displaystyle e^{c_1(y-h_1t)}}+{\displaystyle e^{c_2(y-h_1t)}}$$
$$ K^2_1= 1+a_1{\displaystyle e^{c_1(y-h_1t)}}+a_2{\displaystyle
e^{c_2(y-h_1t)}}$$
$$R_2=K_2=0$$
$m'_y$ is represented in Fig.6 and Fig.7 for different values of the
parameters

\section{Conclusions}

\t The real version of a (2+1) dimensional integrable generalization
of the non-linear Schr\"{o}dinger equation, which has been discussed
by several authors, is studied from the point of view of Painlev\'e
analysis. 
\t In section 2 we  applied the \smm to equation (1.4), obtaining the
leading terms and the singular manifold equations. 
\t In section 3, the Lax pairs were obtained by performing Painlev\'e
analysis on the singular manifold  equations. This  allowed us to
define the eigenfunctions of the Lax pair. 
\t In section 4 we  considered the Lax pairs as system of coupled
PDE's in the fields and eigenfunctions and we obtained the Darboux
transformations between two solutions of (1.4). This permits us to
determine an iterative procedure for obtaining solutions from already
known ones. 
\t Section 6 is devoted to constructing  different kinds of solutions
with solitonic behavior by using the Darboux transformations with
different seminal solutions.

\appendix
\section{Appendix} Substitution of the truncated expansion (2.4) in
(1.4) provides three polynomials in
${1\over \phi}$. Setting each  coefficient of every polynomial at zero
we obtain  the following equations:

\n a) for $m_x+u\omega=0$
$$0=vA+u+A^2\omega\qquad \qquad \qquad \eqno (A.1)$$

\medskip

\n b) for $u_t-u_{xy}-2um_y=0$
$$    0=A_y+qA_x-Aw+vqA+2uq\eqno (A.2)$$
$$ 0=A_t-A_{xy}-A_x(q_x+qv)-vA_y+A(w_x+wv-v_y-vq_x-qv^2-2m_y)
-2u(q_x+vq)\eqno (A.3)$$

\medskip

\n c) for $\omega_t+\omega_{xy}+2\omega m_y=0$
$$    0=A_y+qA_x-Aw+vqA-2\omega qA^2=0\eqno(A.4)$$
$$ 0=-A_t-A_{xy}+2{A_xA_y\over
A}-A_x(q_x+qv)-vA_y+A(w_x+wv+v_y+vq_x+qv^2+2m_y)+ 2\omega
A^2(q_x+\omega q)\eqno(A.5)$$ From (A.2) and (A.4), we can obtain:

$$u=-\left(1\over 2q\right)\left(A_y+qA_x+qvA-wA\right)\eqno (A.6)$$
$$\omega=\left(1\over 2qA^2\right)\left(A_y+qA_x-qvA-wA\right)\eqno
(A.7)$$ (A.6) and (A.7) satisfy (A.1) identically and their
substitution in (A.3) and (A.5) gives (after addition and subtraction):
$$-{A_{xy}\over qA}+{q_xA_y\over q^2A}+{A_xA_y\over qA^2}+{w_x\over
q}-{q_xw\over q^2}=0\eqno (A.8)$$ and
$$A_t-Av_y-{A_xA_y\over A}-2Am_y=0\eqno (A.9)$$ (A.8) can be
integrated in $x$. The result is:
$$\left({1\over q}\right) \left(w-{A_y\over A}\right)+h(y,t)=0\eqno
(A.10)$$ where $h(y,t)$ is the constant with respect to the
integration in $x$. Substitution of (A.10) in (A.6), (A.7) and (A.9)
provides:
$$u=-{A\over2}\left (v+ \left (h+{A_x\over A}\right
)\right)\eqno(A.11)$$
$$\omega= -{1\over {2A}}\left(v-\left(h+{A_x\over
A}\right)\right)\eqno(A.12)$$
$$m_y={1\over 2}\left({A_t\over A}-{A_x\over A}{A_y\over A}-v_y \right
)\eqno(A.13)$$ Finally, imposing the condition that $u$, $\omega$ and
$m$ should be  solutions of (1.4), we obtain:

$$m_x={1\over 4}\left(\left({A_x\over A}+h\right)^2-v^2 \right)\eqno
(A.14)$$ and
$$-A_{xt}+{A_xA_t\over A}+{A_xA_{xy}\over A}+{A_yA_{xx}\over
A}-3{A_x^2A_{y}\over
A^2}+A(v_{xy}-vv_y)+h_yA_x-h_tA+h(A_{xy}-{A_xA_y\over
A})=0\eqno(A.15)$$ Imposing the compatibility condition
$m_{xy}=m_{yx}$ on (A.13) and (A.14), we have:
$$-A_{xt}+{A_xA_t\over A}+{A_xA_{xy}\over A}+{A_yA_{xx}\over
A}-3{A_x^2A_{y}\over
A^2}+A(v_{xy}-vv_y)+h_yA_x+hh_yA+h(A_{xy}-{A_xA_y\over
A})=0\eqno(A.16)$$ Adding and subtracting (A.15) and (A.16), the
result is:

$$h_t+hh_y=0\eqno (A.17)$$
$$\left({A_xA_y\over A^2}-{A_t\over A}\right)_x+\left(v_x-{v^2\over
2}+{1\over 2}\left({A_x\over A}+h\right)^2\right)_y\eqno(A.18)$$

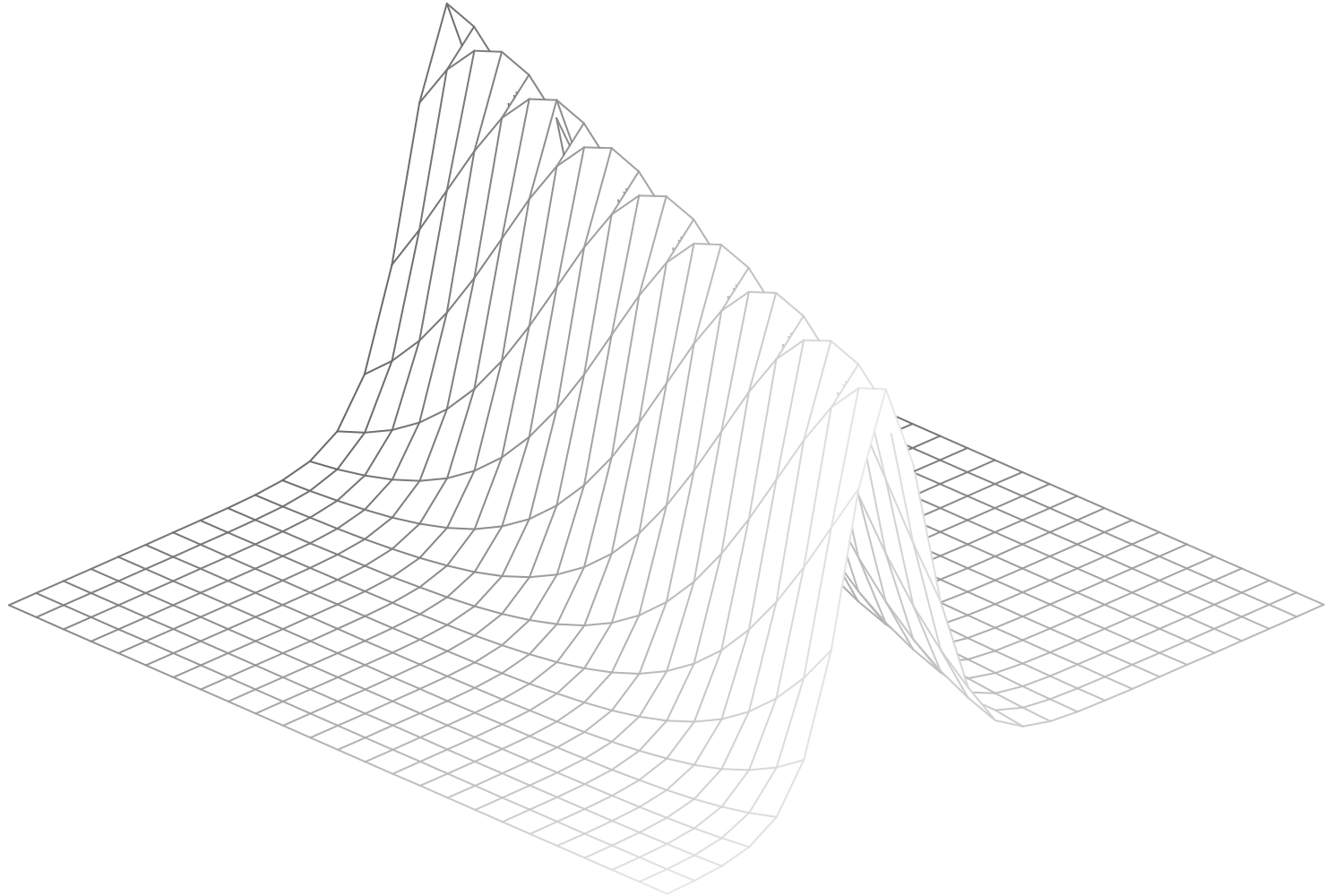
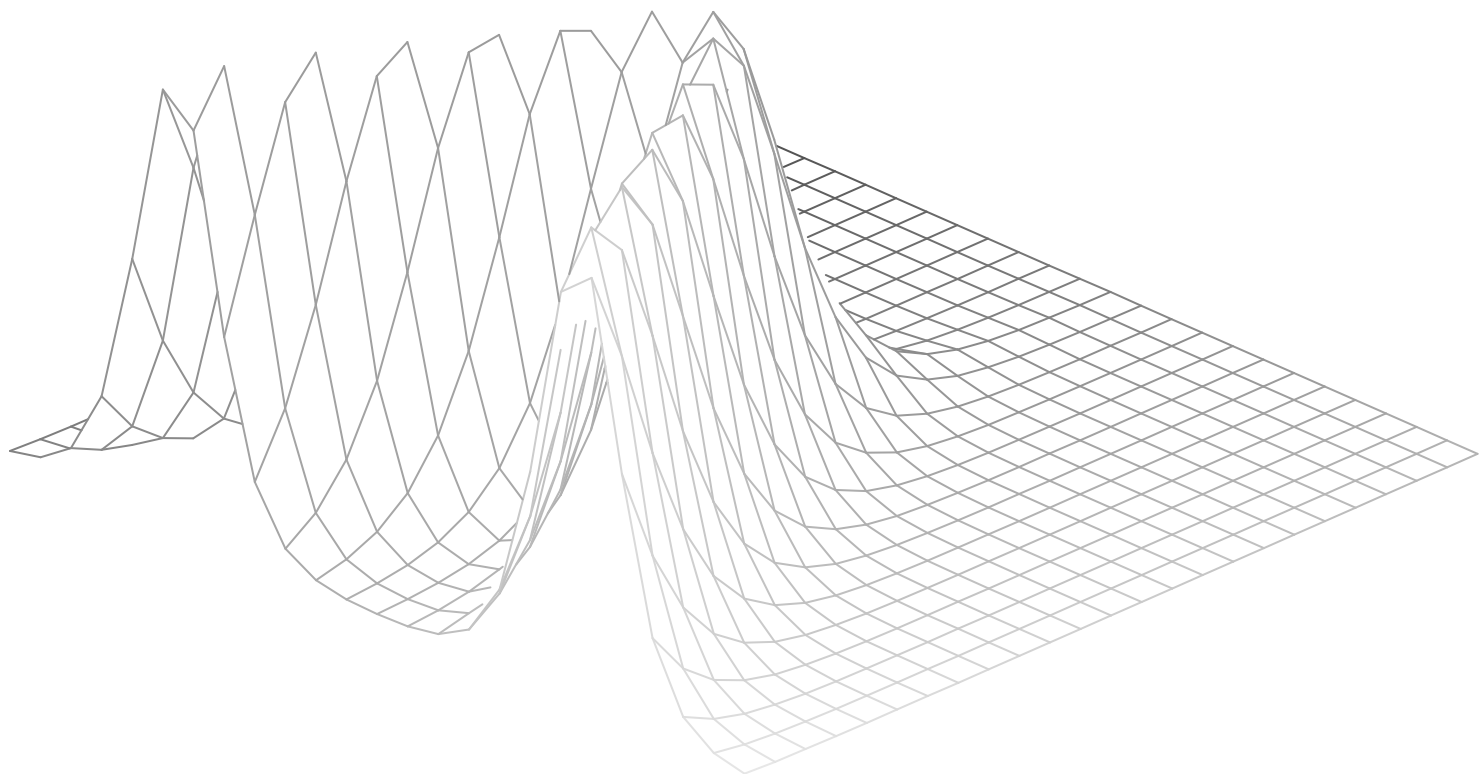
\end{document}